\documentclass[12pt]{article}
\usepackage{amsfonts}
\usepackage{amssymb}
\usepackage{graphicx}
\usepackage{subfigure}
\makeatletter
\input{epsf.tex}
 
\@addtoreset{equation}{section}
\def\section{\@startsection {section}{1}{\z@}{-3.5ex plus -1ex minus
 -.2ex}{1.0ex plus .2ex}{\large\bf}}
\def\subsection{\@startsection{subsection}{2}{\z@}{-3.25ex plus%
 -1ex minus -.2ex}{1.5ex plus .2ex}{\sc}}
 
\advance \voffset by -0.9in
\advance \hoffset by -0.8in
\newcommand{\ZZ}{\mathbb{Z}}

\newcommand{\RR}{\mathbb{R}}
\newcommand{\CC}{\mathbb{C}}

\def\bea{\begin{eqnarray}}
\def\eea{\end{eqnarray}}

\begin{document}
\baselineskip 18pt
\parskip 7pt
\begin{flushright}
nlin.PS/0201047 \\
HWM01-44
\end{flushright}
\vspace{0.7cm}

\begin{center}
\baselineskip 24 pt
{\LARGE Unstable manifolds and Schr\"odinger dynamics}\\
{\LARGE  of Ginzburg-Landau vortices}

\vspace{1cm}

{\Large O.~Lange\footnote{address since Dec. 2001: 
Max-Planck-Institute for Biophysical Chemistry,
Theoretical Molecular Biology Group,
D-37077 G\"ottingen, Germany;
 e-mail {\tt oliver.lange@mpi-bpc.mpg.de}}
and B.~J.~Schroers\footnote{e-mail {\tt bernd@ma.hw.ac.uk}}  \\
Department of Mathematics, Heriot-Watt University\\
Edinburgh EH14 4AS, United Kingdom}

\baselineskip 18pt

\vspace{0.1cm}

January 2002

\vspace{0.3cm}

\end{center}

\begin{abstract}
\noindent 
The time evolution of  several  interacting Ginzburg-Landau 
vortices according to an equation of Schr\"odinger type is 
approximated by motion on a  finite-dimensional manifold. 
That manifold is defined as an unstable manifold of an auxiliary 
dynamical system, namely the gradient flow of the Ginzburg-Landau 
energy functional. For two vortices the relevant unstable manifold 
is constructed  numerically and the induced dynamics is computed.  
The resulting model provides a complete picture of  the vortex 
motion for arbitrary vortex separation, including well-separated and 
nearly coincident vortices.

\end{abstract}

\centerline{ AMS classification scheme numbers: 35Q55, 37K05, 70K99}

\section{Ginzburg-Landau vortices and their dynamics}

Vortices play a fundamental role in a large variety of physical systems,
ranging from ordinary fluids over condensed matter to the early universe.
This variety is reflected in the mathematical models used to describe
the formation, structure and dynamics of vortices. In fluid dynamics,
the basic ingredient of the mathematical model is the velocity field
of the fluid, and the vortex is a particular configuration of that 
velocity field. In  condensed matter theory, the mathematical models are 
field theories. The basic field of such models is a complex valued 
scalar field, and vortices are particular configurations of that field.
One important difference between vortices in ordinary fluids and 
those in condensed matter systems is that the total
vorticity can take 
arbitrary real values in the former but is quantised in the latter. 
Nonetheless, it is sometimes 
 possible to establish a precise mathematical connection
between field theory models and those used in ordinary fluid dynamics.

In this paper we are concerned with vortices in the Ginzburg-Landau model.
For general background and references we refer the reader to the book
\cite{BBH}.
We are interested in the dynamics of such vortices in the plane, so the 
basic field of the model is the function
\bea
\psi: \RR\times \RR^2 \rightarrow \CC,
\eea
 depending on  time $t\in \RR$ and 
 spatial coordinates $x=(x_1,x_2)\in \RR^2$. 
Sometimes we also use polar coordinates $(r,\theta)$ in the plane
and  parametrise the field $\psi$ in terms of its argument and modulus
via $\psi = \sqrt\rho\exp(i\phi)$, where $\rho=|\psi|^2 $.
In this paper $\psi$ will be used to denote both time-dependent and 
static configurations. For static configurations we often  suppress
the first argument and simply write $\psi(x)$. For 
configurations varying with time we sometimes write $\dot{\psi}$ for 
the partial derivative $\partial_t\psi$.
Often we write $\partial_1$ for $\partial /\partial x_1$, $\partial_t$
for $\partial/\partial t$ and so on.

The Ginzburg-Landau
energy is the following functional of $\psi$:
\bea
\label{GLenergy}
E[\psi] = \frac 12 \int  {\nabla\psi}\nabla\bar\psi + 
\frac 12( |\psi|^2-1)^2 \, d^2x.
\eea
We impose the boundary condition
\bea 
\label{boundary}
\lim_{r\rightarrow \infty}|\psi (r,\theta)|=1,
\eea
which implies that  for sufficiently large $R$ the field
\bea
\psi_R(\theta)=\frac {\psi } {|\psi  |} (R,\theta)
\eea
is  a well-defined  map from the spatial circle 
of large radius $R$  to the unit circle in $\CC$.
It therefore has an associated integer winding number or degree,
which can be computed via the following line integral over the 
circle $S^1_R$ of radius $R$
\bea
\label{degform}
\mbox{deg}[\psi]=\frac 1 {2\pi i}\oint_{S^1_R}d\ln \psi.
\eea
This integer, denoted $n$ in the following, is also  called the vortex number.

The variational derivative of the Ginzburg-Landau 
functional (\ref{GLenergy}) with respect to $\bar\psi$ is
\bea 
\label{functionalder}
\frac{\delta E}{\delta \bar{\psi}} = -\frac 1 2 
\left(\Delta \psi + (1- |\psi|^2 )\psi\right)
\eea
and,  because of the reality of $E$,
\bea
 \frac{\delta E}{\delta {\psi}}=\overline{\frac{\delta E}{\delta \bar{\psi}}}.
\eea
The  Ginzburg-Landau  equation is the equation for critical points of the
Ginzburg-Landau energy, so it reads
\bea
\label{LGequation}
-\Delta \psi + ( |\psi|^2 -1)\psi =0.
\eea
A Ginzburg-Landau
vortex is a solution of this equation
with non-vanishing vortex number $n$. 
It is shown in \cite{OS3} that for any configuration of non-vanishing
vortex number which attains the limit (\ref{boundary}) uniformly in $\theta$
the Ginzburg-Landau energy is necessarily infinite. 
 The origin of this divergence is not difficult to understand and 
also explained in \cite{OS3}. The essential point is that 
the gradient terms  in the energy density contain the term
$|\partial_\theta\psi|^2/r^2$ which, for $n\neq 0$,
leads to a logarithmic divergence. Since the divergence only depends
on $n$ and not on any other details of the configuration, it can
be removed by introducing a smooth  cut-off function
\bea
\chi_R(x)=\left\{\begin{array}{cl}      0 & \,\,\mbox{for}\,\, |x|\leq R\,\,\\
     1 &\,\,\mbox{for} \,\, |x|\geq R+R^{-1}  \end{array}
\right.            
\eea
and defining the renormalised energy functional
\bea
E_{\mbox{\tiny ren}}[\psi]= \frac 12 \int  {\nabla\psi}\nabla\bar\psi -
 \frac{ \mbox{deg}[\psi]^2}{r^2} \chi_R +
\frac 12( |\psi|^2-1)^2 \, d^2x.   
\eea
This renormalisation procedure 
is natural both from the point of view of physics and in numerical
investigations. The point is that  the total vortex number $n$ 
is conserved during the  time evolution. When studying the interacting
dynamics of $n$ vortices during a finite time interval we can choose 
$R$ so large that all the vortices remain well inside
 the disc of radius $R$  during that time interval.  
 The  renormalisation procedure removes a divergence which only
depends on the conserved quantity $n$ but not on any other details
of the dynamics.

In studying the dynamics of vortices one 
has a choice of several different ways of extending the static
Ginzburg-Landau  equation to 
a time-dependent evolution equation. We are interested in the 
following first-order  equation of Schr\"odinger type, often
called the Gross-Pitaevski equation:
\bea
\label{GPequation}  
i \frac{\partial \psi }{\partial t} = -\Delta \psi + ( |\psi|^2 -1)\psi.  
\eea
The vortex dynamics dictated by this equation
has been studied in a large number of publications. In two seminal
papers \cite{Neu1,Neu2} 
Neu showed that in a scaling limit where the vortex
size shrinks to zero, the time evolution according to the {\it partial} 
differential equation (\ref{GPequation}) reduces to a set of 
coupled  {\it ordinary}
 differential equation for  the centres of vorticity. Moreover, 
he showed that set of ordinary equations to be the  
Kirchhoff-Onsager law for  the motion of fluid vortices in 
incompressible, nonviscous
 two-dimensional flows. Neu's work has in turn inspired a number
of authors. 
Some, mathematically motivated,  have investigated the scaling limit
further and have put his work on a more
rigorous mathematical footing, see  \cite{LX1,LX2} and \cite{CJ}.  
Others, starting
with the interpretation of Neu's work as a 
 finite-dimensional  approximation to 
the infinite-dimensional dynamical system governed by the Gross-Pitaevski
equation, have tried to go beyond this approximation. 
Physically, one may think of Neu's approximation as a point-particle
approximation to vortex dynamics in a field theory. This approximation
is expected to be reasonable when the vortices are well-separated 
and moving slowly. However, when the vortices overlap, the point-particle
approximation is poor. Similarly, when the vortices move rapidly we 
expect them to excite radiation in the field theory which is not 
captured by the point-particle approximation.

More recently, Ovchinnikov and Sigal have derived Neu's approximation
from a different point of view and have computed leading radiative 
corrections to it in a series of papers \cite{OS1,OS2}.
They work with the  Lagrangian from which the Gross-Pitaevski 
equation can be derived and 
explicitly study 
the truncation of the infinite dimensional Gross-Pitaevski dynamical
system  to a finite-dimensional family of multi-vortex
configurations with pinned centres of vorticity. The induced Lagrangian
on that finite dimensional family reproduces the Kirchoff-Onsager
law when the vortex centres are well-separated. Moreover, the approach
makes it possible to study the coupling between the vortex motion and 
radiative modes and to compute radiative corrections. However, Ovchinnikov
and Sigal's family of pinned  multi-vortex configurations is less useful
when trying to understand the dynamics of overlapping vortices. When vortices
get close together pinning the vortex centres is mathematically awkward 
and physically unnatural. In this paper we  study the truncation of the 
Gross-Pitaevski dynamical system to a finite dimensional dynamical
system  using a 
different set of multi-vortex configurations. Our configurations 
are similar to 
 those of Ovchinnikov and Sigal when the vortices are well-separated,
but we believe they  provide a  more
accurate description of vortex dynamics when the vortices overlap. 
Our approach is inspired by an approximation scheme 
proposed by Manton \cite{Manton1} 
in the context of Lagrangian soliton dynamics.
Manton considered the problem of defining a smooth
finite-dimensional  family of multi-soliton configurations which
could be used as the configuration space
for  a  finite-dimensional low-energy approximation
to multi-soliton dynamics. His proposal is to consider 
 an auxiliary evolution equation, namely
the gradient flow in the potential  energy functional,
and to use the unstable manifold of a suitable saddle point
as  the truncated configuration space.
 Such an  unstable manifold is the union of paths of steepest
descent from the saddle point. 
In practice this scheme is not easy to implement, and so far 
it has only been used to study soliton dynamics in one spatial dimension
\cite{MM}. In this paper we show that it is very well suited to studying 
the dynamics of overlapping Ginzburg-Landau
vortices. The basic reason is  
that in the Ginzburg-Landau model there are well-known rotationally
symmetric saddle point solutions for all vortex numbers $|n|>1$
which can be used for the construction of the unstable manifold. 
We focus on the case $n=2$ and find that the relevant unstable manifold is
two-dimensional. We describe its geometry, compute the induced
Lagrangian  and use it to study the relative motion of two vortices
at arbitrary separation.

\section{The Gross-Pitaevski equation and its symmetries}

For our study of  the Gross-Pitaevski equation it is essential
that we can derive it from a Lagrangian. 
The required Lagrangian is the following functional of 
 time-dependent fields $\psi$
\bea
\label{GPlag}
L[\psi]=T[\psi,\dot\psi]-E[\psi],
\eea
where the kinetic energy functional $T$  is given by
\bea 
 T(\psi,\dot\psi)=-\frac 12 \int \mbox{Im}( \bar\psi\dot\psi)\, d^2x
\eea
and the potential energy functional $E$  is the Ginzburg-Landau
functional (\ref{GLenergy}). One checks that, essentially because of the 
linearity of the kinetic energy in  $\dot \psi$, the total energy or 
Hamiltonian is equal to the potential energy $E$. Using the 
formula (\ref{functionalder}), the Euler-Lagrange
equation of $L$ 
\bea
\partial_t\psi=-2i\frac{\delta E}{\delta \bar \psi}, 
\eea
is readily seen to be 
 the Gross-Pitaevski equation (\ref{GPequation}). 

We define the configuration space to be  the space of smooth
fields $\psi$ satisfying the boundary
condition (\ref{boundary}), i.e.
\bea
\label{confspace}
{\cal C}=\{\psi :\RR^2\rightarrow \CC | \lim_{|x|\rightarrow \infty }
|\psi(x)|  =1\}.
\eea
As mentioned earlier, 
the boundary condition means that  
every  $\psi\in {\cal C}$ has an associated 
integer winding number  or degree. Using Stokes's theorem and taking 
the radius $R$ to infinity in (\ref{degform}) one shows that 
 the degree can also  be written as the integral
\bea
\mbox{deg}[\psi]=\int_{\RR^2} \gamma,
\eea
where the integrand
\bea
\label{vorticity}
\gamma =\frac{1}{2\pi i} d\bar\psi\wedge d\psi
\eea
is called  the vorticity.
The integer degree cannot change under the continuous time evolution,
which can therefore be restricted to one of the topological sectors
\bea
{\cal C}_n=\{\psi \in {\cal C}|\,\mbox{deg}[\psi]=n\}.
\eea
For each $n$, the pair $({\cal C}_n, L)$  is an 
infinite dimensional  Lagrangian dynamical system.
{}From the Lagrangian viewpoint ${\cal C}_n$ is the configuration
space and a motion is a path $t\mapsto \psi(t,\cdot)$  in ${\cal C}_n$ 
which satisfies the evolution equation (\ref{GPequation}). 
Alternatively, we can adopt the Hamiltonian point of view.
Then ${\cal C}_n$ should be thought of as the phase space and 
the kinetic part of the Lagrangian (\ref{GPlag}) defines a 
symplectic structure on ${\cal C}_n$. This is a two-form $\Omega$ 
on the tangent space of ${\cal C}_n$. For a more careful discussion
of that tangent space we refer the reader to \cite{OS3}, but
for our purpose it is sufficient to think of tangent vectors
as complex valued functions which tend to zero at infinity.
If $\xi$ is such a tangent vector we  use the  notation
introduced in \cite{OS3} for elements of the complexified 
tangent space:
\bea
\label{vectnot}
\vec{\xi} = \pmatrix{\xi \cr \bar{\xi}}.
\eea
If  $\xi$ and $\eta$ are  two
elements of the same  tangent space the symplectic form is defined via
\bea
\label{symform}
\Omega(\vec{\xi},\vec{\eta})=
\frac {1}{2i} \int (\xi\bar{\eta}-\bar{\xi}\eta)\, d^2x.
\eea
It follows that the Poisson bracket of 
$\psi$ and $\bar{\psi}$   is
\bea
\label{poisson}
\{ \psi,\bar \psi\}=2i.
\eea
One checks that the Gross-Pitaevski equation can then be 
written in the canonical form
\bea
\dot\psi =\{E[\psi],\psi\}.
\eea
As an aside we point out that the total vorticity or degree is 
related to the pull-back of the symplectic form $\Omega$ 
evaluated on the vector fields $\partial_1$ and 
$\partial_2$ on $\RR^2$ : 
\bea
\psi^*\Omega(\partial_1,\partial_2)
 =-\pi  \int_{\RR^2}\gamma.
\eea
Thus if $\psi \in {\cal C}_n$ then $\psi^*\Omega(\partial_1,\partial_2)
= -n\pi$.

The Lagrangian $L$ has a large  invariance group. Writing $R\in SO(2)$ 
for spatial rotations and $d\in \RR^2$ for translations in the plane,
 elements $(R,d)$ of the  Euclidean group
\bea
E_2= SO(2)\ltimes \RR^2
\eea
act on fields via pull-back
\bea
\psi\mapsto \psi\circ (R,d)^{-1},
\eea
i.e. $\psi \circ (R,d)^{-1}(x)=\psi(R^{-1}(x-d))$. 
Both the Lagrangian and 
the degree are invariant under this action. Similarly, unit complex
numbers $e^{i\alpha}\in U(1)$ act on fields $\psi$ as phase rotations
\bea
\psi\mapsto e^{i\alpha}\psi
\eea
and leave both the Lagrangian and the degree invariant.
Spatial reflections about any axis leave the Lagrangian invariant
but change the sign of the degree. The same is true for the combination
of  complex conjugation with time reversal $T:t\mapsto -t$ .
 However, with $S:(x_1,x_2)\mapsto (x_1,-x_2)$,   the combination
\bea
\label{refsym}
C:\psi\mapsto \bar\psi\circ (ST)
\eea
leaves the degree and the Lagrangian invariant. 
To sum up, for each $n \in \ZZ$
the group 
\bea
\label{symgroup}
C\ltimes \left(E_2\times U(1)\right)
\eea
acts on ${\cal C}_n$ preserving the Lagrangian $L$ and hence the 
equations of motion.
The two-dimensional Galilean  group provides an additional more subtle
symmetry. If $\psi(t,x)$ solves the Gross-Pitaevski equation, then 
so does the  configuration
\bea
\label{galileo}
\psi_v(t,x)=e^{i(\frac 1 2 x\cdot v -\frac 1 4 v^2t)}\psi(t,x-vt),
\eea
where the parameter $v \in \RR^2$ is physically interpreted as
the boost velocity.

To end this section we note
 the conservation laws which follow from the symmetry group (\ref{symgroup}). The Noether charge which is conserved due
to  the  invariance under phase rotations is 
\bea 
\label{qcharge}
Q[\psi]=\int_{\RR^2}|\psi|^2 \,d^2x.
\eea
Invariance under spatial rotations leads to the conserved charge
\bea
\label{jcharge}
J[\psi]=\int_{\RR^2} \mbox{Im}(\bar\psi\partial_\theta\psi) \, d^2x
\eea
and invariance under translations leads to the conservation of 
the vector $(P_1,P_2)$ with components
\bea
P_i[\psi]=\int_{\RR^2} \mbox{Im}(\bar\psi\partial_i\psi) \, d^2x \quad i=1,2.  
\eea
All of the above charges have to be handled with care because 
 the integrals defining them do not generally converge
for configurations with non-vanishing degree.  
In  \cite{PT} a related field theory
was studied and it was pointed out  
that the Noether charges can be related to moments of the 
vorticity. In our cases the moments
\bea
\label{Jmoment}
\tilde J[\psi] =-\pi\int_{\RR^2} (x_1^2+x_2^2) \gamma
\eea
and 
\bea
\label{Pmoment}
\tilde P_1[\psi] =2\pi \int_{\RR^2} x_2\gamma \quad \mbox{and}
\quad  \tilde P_2[\psi] =-2\pi \int_{\RR^2} x_1\gamma 
\eea
are also conserved during time evolution according to (\ref{GPequation})
and can be obtained from  the Noether charges $J$, $P_1$
and $P_2$  by integration
by parts. For some  configurations the integrals defining $\tilde J$ and 
$\tilde P_i$ are convergent even when those defining $J$ and $P$ are not.
The  Galilean  symmetry also implies a conservation law,
but it will not be required  in this paper.

\section{The $n=2$ saddle point and its unstable mode}

Imposing extra symmetry is the easiest way of finding static
solutions of the Ginzburg-Landau equations. The largest invariance group
one can impose on configurations of degree $n$ is  a group
containing reflections and combinations of  spatial rotations $R(\chi)$ 
by an angle $\chi$ with suitable phase rotations. More precisely we define
the subgroup
\bea
\label{sadsym}
R_n=C\ltimes 
\{ (R(\chi),e^{in\chi}) \in SO(2)\times U(1)|\chi\in [0,2\pi)\}
\eea
of the symmetry group (\ref{symgroup}). Configurations invariant under
this group are of the form
\bea
\psi_n(r,\theta)=f_n(r)e^{in\theta},
\eea
where $f_n$ is real and  satisfies 
the boundary condition 
\bea
\label{unique}
f_n(0)=0 \quad  \mbox{for}\quad  n\neq 0 \quad
\mbox{and} \lim_{r\rightarrow \infty}f_n(r)=1.
\eea
The Ginzburg-Landau equation implies the following ordinary differential
equation for $f_n$
\bea
\frac{1}{r}\frac{d}{dr}\left(r\frac{df_n}{dr}\right) -\frac{n^2}{r^2}f_n
+(1-f_n^2)f_n=0.
\eea
As proved for example in \cite{BBH},
 this equation has a unique solution satisfying the 
boundary conditions (\ref{unique}) for each $n$,
 and in the following we use $f_n$ to 
denote that solution. Near $r=0$ and  for large $r$ the equation can
be solved approximately
\bea
f_n(r)&\sim& A_n r^n \quad \mbox{for small}\,r, \nonumber \\
f_n(r)&\sim& 1-\frac{n^2}{2r^2} \quad \mbox{for large}\, r,
\eea
where $A_n$ are real constants. 
However,  no exact expression for 
$f_n$ in terms of standard functions is known.
For $|n|=1$ the resulting vortex configuration is a stable solution
of the Ginzburg-Landau equation. For $|n|> 1 $ the solution is an unstable 
saddle point. Bounds on the number of unstable modes were
  derived by Ovchinnikov and Sigal  in \cite{OS3}.  
It follows from their analysis that for 
$n=2$ there is  precisely one unstable mode.
Since this mode plays an important role in our analysis, we describe
and compute it explicitly. For further general
background we refer the reader to 
\cite{OS3}. The required unstable mode is an eigenvector
 of the Hessian of the Ginzburg-Landau energy
 \bea
\mbox{Hess}\,E[\psi]=\pmatrix{\frac{\delta^2 E}{\delta \psi\delta \bar{\psi}}
&\frac{\delta^2 E}{\delta \bar{\psi}^2}\cr
\frac{\delta^2 E}{\delta \psi^2} &
\frac{\delta^2 E}{\delta \psi\delta \bar{\psi}}
}.
\eea
For the saddle point $\psi_2$ the Hessian takes the form
\bea
\mbox{Hess}\,E[\psi_2]= \frac 1 2\pmatrix{-\Delta +2\rho_2-1
&\rho_2e^{4i\theta}\cr
\rho_2e^{-4i\theta}&
-\Delta +2\rho_2-1
},
\eea
where $\rho_2=|\psi_2|^2$.
In terms of the  notation (\ref{vectnot})
 the  eigenvalue equation
\bea 
\label{hesseigen}
2\mbox{Hess}\,E[\psi_2]\vec{\xi}=\lambda
\vec{\xi}
\eea
is thus equivalent to
\bea
\label{fluctuate}
-\Delta \xi + (2\rho_2-1)\xi +\rho_2 e^{4i\theta}
\bar \xi = \lambda\xi,
\eea
which can also be obtained by linearising the Ginzburg-Landau
equation around the saddle point $\psi_2$.
If $\xi$ solves (\ref{fluctuate}) then the  leading term 
in the difference   $E[\psi_2+\xi] - E[\psi_2]$
is the second order term
\bea 
E^{(2)}(\vec{\xi})&=&\frac 1 2 
\int (\vec{\xi})^\dagger\, \mbox{Hess}\,E[\psi_2]\,\vec{\xi}
\nonumber \\ 
&=&\frac{1}{2} \lambda \langle\vec{\xi}, \vec{\xi}\rangle,
\eea
where we used the inner product
\bea
\label{metric}
\langle\vec{\xi}, \vec{\eta}\rangle =\frac 1 2 \int(
\xi\bar{\eta} +\eta\bar{\xi})\,d^2x.
\eea

To study the  eigenvalue problem (\ref{fluctuate}) and to compute
the variation in the energy 
 it is best  to expand the 
$\theta$-dependence of $\xi$
into a  Fourier series. 
 The term $\rho_2 e^{4i\theta}$ in the equation leads to a coupling 
between Fourier modes $e^{ik\theta}$ with $k$ differing by $4$,
but all other modes decouple. As explained in \cite{Dziarmaga},    
the unique eigenfunction with a  negative
eigenvalue is of the form 
\bea
\label{eigenfun}
\xi(r,\theta) =u(r) + v(r)e^{4i\theta}.
\eea
Inserting this expression into (\ref{fluctuate})
leads to  coupled equations for the radial functions 
$u$ and $v$
\bea
\label{coupled}
    -\frac 1 r \frac{d}{dr}\left(r\frac{ d u}{d r}\right)
 + \left(2\rho_2 -1\right)u + \rho_2 v &=&\lambda u \nonumber \\ 
  -\frac 1 r \frac{d} {dr} \left(r \frac{dv}{d r}\right)
+ \left(\frac {16} {r^2} + 2\rho_2 - 1\right)v  +\rho_2 u &=&\lambda v. 
\eea
In terms of  $u$ and $v$ 
the second variation of the Ginzburg-Landau
energy for a fluctuation  $\xi$ of the form (\ref{eigenfun})
about the stationary point $\psi_2$ is
\bea
\label{energyvar}
E^{(2)}(\vec{\xi}) = \pi\int_0^\infty  \,\left[ \left(\frac {du }{dr}\right)^2
+ \left(\frac {dv }{dr}\right)^2 + (2\rho_2-1)(u^2+v^2) + 2\rho_2 uv
+\frac{16}{r^2}v^2\right]\,r dr,
\eea
where we have assumed regularity of $u$ and $v$ at the origin 
and exponential decay at infinity in the integration by parts.
This assumption will be justified below. 
Note that, in terms of the functions $u$ and $v$, 
\bea
\langle\vec{ \xi}, \vec{\xi}\rangle =
2\pi\int_0^\infty \left( u^2 + v^2\right)\,\,r dr.
\eea

The eigenvalue (\ref{coupled})
was first 
 studied numerically by Dziarmaga in \cite{Dziarmaga},
using a shooting method. Near $r=0$ the leading terms for a regular
solution are
\bea
u(r)\sim 1-\frac{1+\lambda}{4}r^2  \quad \mbox{and}\quad
v(r)\sim a r^4, 
\eea
where $a$ is an undetermined parameter. For large $r$, it is convenient
to write the equation in terms of the linear combinations $u_+
=u+v$ and $u_-=u-v$:
\bea
\label{pmcoupled}
    -\frac 1 r \frac{d}{dr}\left( r\frac{ du_+}{d r} \right) 
 + \left(\frac {8} {r^2}+ 3\rho -1\right)u_+ \frac{8}{r^2}u_-&=& \lambda u_+\\
    -\frac 1 r \frac{d} {dr} \left( r \frac{du_-}{d r}\right) 
+ \left(\frac {8} {r^2} + \rho - 1\right) u_- +  \frac{8}{r^2}u_+& =&  \lambda u_-.
\eea
Now the equations decouple for large $r$ and one deduces the asymptotic
form
\bea
u_+&\sim& A\frac{\exp(-\sqrt{|\lambda| +2}\,r) }{\sqrt{r}} +
 B\frac{\exp(\sqrt{|\lambda|+2}\,r)}{\sqrt{r}} \nonumber \\
u_-&\sim& C\frac{\exp(-\sqrt{|\lambda|}r)}{\sqrt{r}} +
 D\frac{\exp(\sqrt{|\lambda| }r)}{\sqrt{r}}.
\eea
Solving the eigenvalue problem with a shooting method means
determining the two parameters $a$ and $\lambda$ so that  
the solutions $u_+$ and $u_-$ both 
 decay exponentially for large $r$ i.e.  $B=D=0$.
Such a parameter search in a two-dimensional parameter space 
 is tricky. We start our search 
by assuming that $v$ is negligible compared to $u$ and take 
the trial function $\xi_0(r,\theta)=u_0(r)$
 where
\bea
u_0(r) =\mbox{sech}\left( \frac{3}{4} r\right),\
\eea
One then finds that 
\bea
 E^{(2)}(\vec{\xi_0}) = -0.378\times  \frac 1 2 \langle
 \vec{\xi}_0,\vec{\xi}_0\rangle,
\eea
showing that 
$\lambda \leq -0.378$ and hence that the value of $-0.168$
for  the eigenvalue given in 
\cite{Dziarmaga} is incorrect. In our search, starting with $\xi_0$,
we find the eigenvalue
\bea
\lambda = -0.41869,
\eea
with eigenfunctions $u$ and $v$ shown in figure 1.
Note that $u$ 
is qualitatively very similar to $u_0$. However, $v$ is non-vanishing
and negative.  This is not surprising.
Looking at the energy expression (\ref{energyvar}) we note 
that it is energetically favourable for the functions $u$ and $v$
 to have opposite signs. 
\begin{figure}[htb]
\epsfxsize=3 in
\bigskip
\centerline{\epsffile{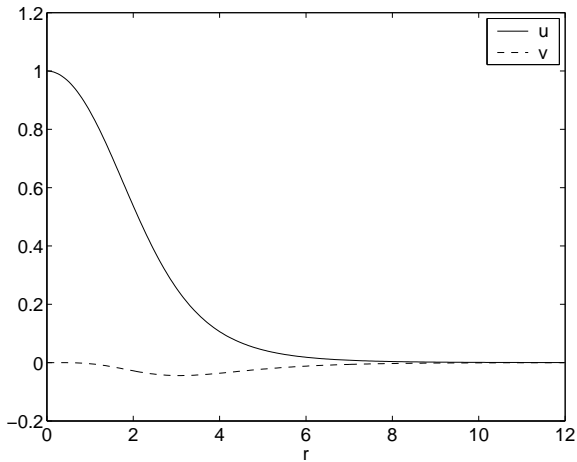}}
\caption{\baselineskip=12pt {\it
The unique negative eigenmode of the Hessian at
 the saddle point $\psi_2$. The plot shows the radial functions in 
the parametrisation (\ref{eigenfun}) .}}
\bigskip
\label{fig1}
\end{figure}

\section{Unstable manifold and truncated dynamics}

 The unstable 
manifold of a saddle point is the union of paths of steepest descent
from the saddle point. To define paths of steepest descent, one
requires both a metric and a potential. 
Suppose $x^i$ are local coordinates on a finite-dimensional manifold with 
 potential energy $V$ and the metric is represented by the matrix $g_{ij}$. 
Then the paths of steepest descent are 
solutions to the  gradient flow equations 
\bea
g_{ij}\frac{d x^j}{d\tau} =-\frac{\partial V}{\partial x^i},
\eea
with $x^i(\tau)$ approaching the relevant saddle point as $\tau \rightarrow
-\infty$. It is important to keep track of the metric even when it is flat.
For example, gradient flow on $\RR^2$ with the standard 
flat metric takes the 
following form in complex coordinates $z=x_1+ix_2$:
\bea
\frac{d z}{d\tau}=-2\frac{\partial V}{\partial \bar z}. 
\eea

In the field theory we are considering here the potential energy
is the Ginzburg-Landau
potential  energy functional (\ref{GLenergy}) and the metric is 
the flat  metric. Since we work with complex-valued fields, 
tangent vectors are also complex-valued functions and the metric
is the inner product of tangent vectors defined in (\ref{metric}).
The gradient flow equation is the following non-linear heat equation
\bea
\label{heat}
\partial_{\tau} \psi =-2 \frac{\delta E}
{\delta\bar  \psi} = \Delta \psi + (1- |\psi|^2)\psi. 
\eea
Here $\tau$ is an auxiliary ``time'' variable, which we distinguish
notationally from the time variable  $t$ used in the Gross-Pitaevski
equation.

We have solved this equation numerically on a grid of size $100\times 100$,
corresponding physically to the  square $[-20,20]\times [-20,20]$ in the 
$x_1x_2$-plane. To generate the gradient flow curve we start with
the saddle point configuration $\psi_2$ and add a small perturbation
$\delta \psi = \epsilon \xi$, where $\xi $ is the negative eigenmode
(\ref{eigenfun}) found in the previous section. At the boundary
of the lattice we impose the Dirichlet boundary condition 
$\psi(r,\theta)=e^{2i\theta}$.  The resulting gradient flow  curve   
does not depend significantly  on $\epsilon$ provided
it is small enough.  When $\epsilon =0$, the discretisation effects
provide sufficient perturbation to make sure that the gradient flow 
curve moves away from the saddle point $\psi_2$, but the initial
flow is very slow.
It is therefore numerically convenient to use a non-vanishing 
value of $\epsilon$. In figures  2 and 3  
we show snapshots of the field and of the 
energy density during the early
 stage of the gradient flow, generated by starting
with $\delta \psi = 0.0005\, \xi$. During the gradient flow, the 
unstable $n=2$ vortex splits into two $n=1$ vortices which drift apart.
The saddle point solution $\psi_2$ has an energy density which has
a degenerate maximum  on a ring. The splitting process begins with the 
development of two local maxima, as shown in configuration A in figure
3. As the splitting process continues,
these local maxima become two distinct vortices. Note,
however, that  during the early phase of the splitting process, the vortices
overlap and deform each other significantly. This is clearly visible
in configuration B in figure 3. Only in the later stages of the 
splitting process do the vortices  resemble two standard
 $n=1$ vortex solutions.

\begin{figure}[htb]
\epsfxsize=2.3 in
\bigskip
\centering
        \begin{minipage}[b]{0.5\textwidth}
        \centering
            \subfigure{
              \includegraphics[width = 2.3 in]{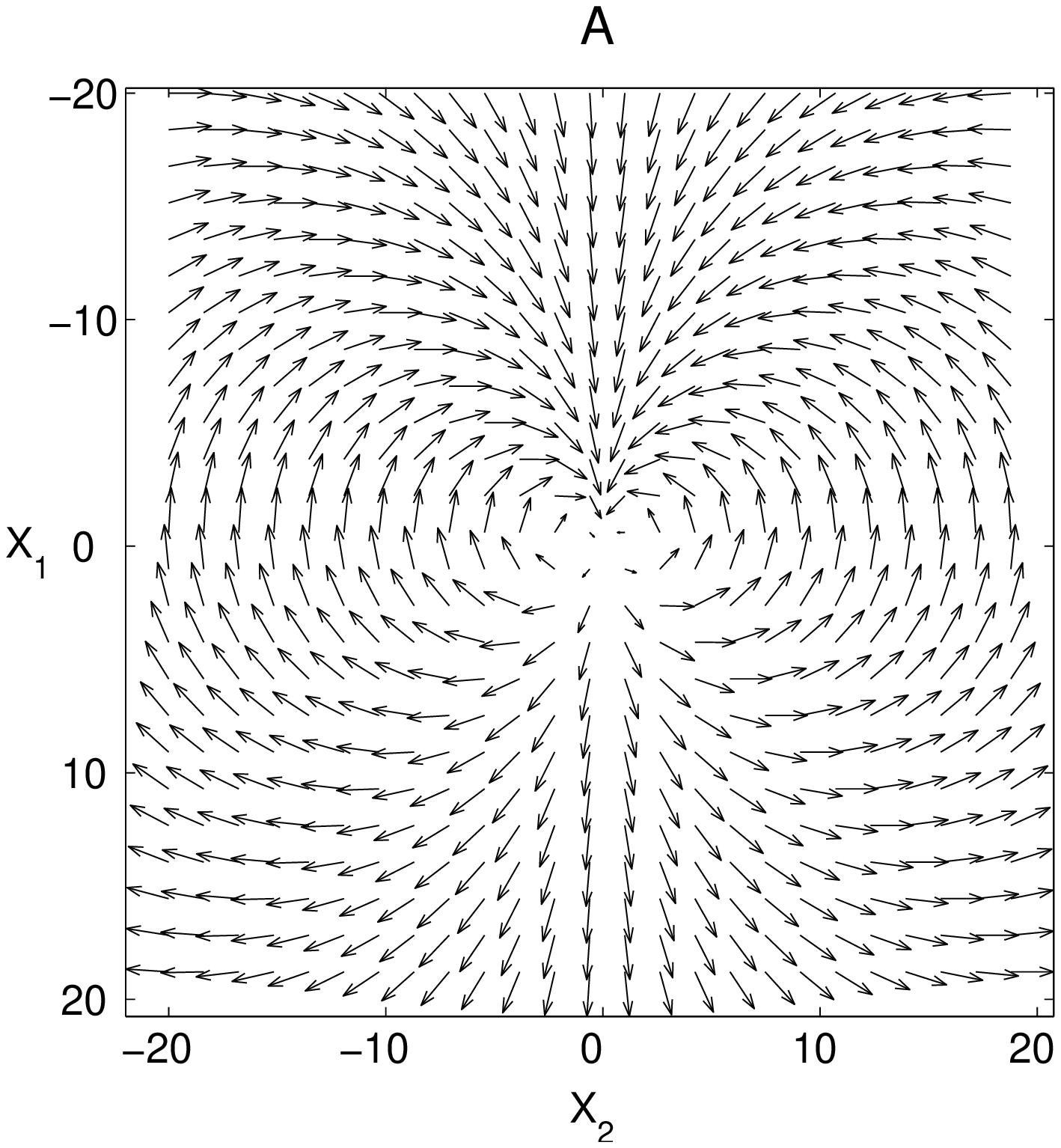}}
        \end{minipage}%
        \begin{minipage}[b]{0.5\textwidth}
        \centering
            \subfigure{
             \includegraphics[width = 2.3 in]{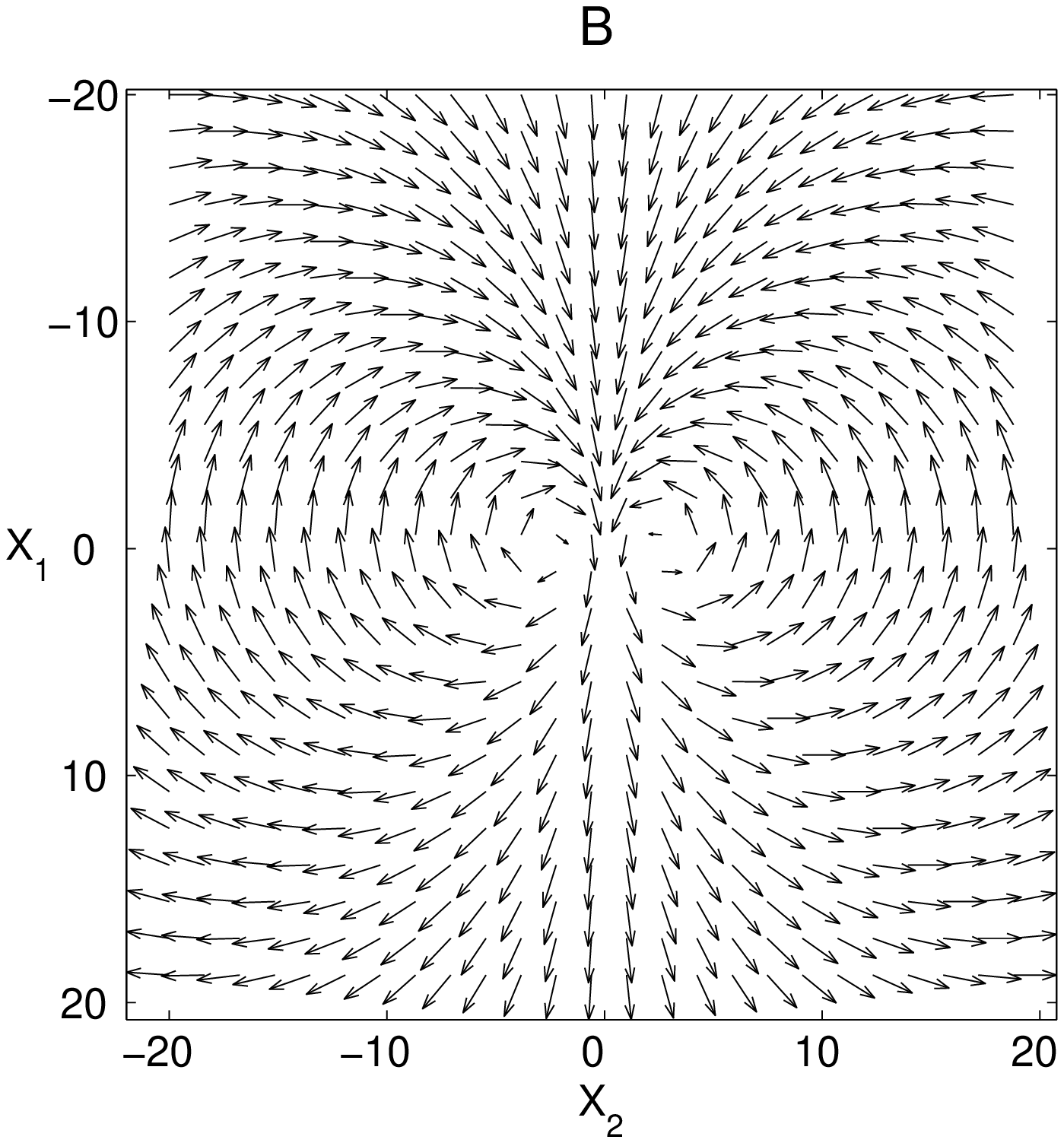}}
        \end{minipage}
\caption{\baselineskip=12pt {\it
 Vortex configurations in the $x_1x_2$-plane during
          the early stages of the gradient flow from the saddle point 
$\psi_2$.}
}
\bigskip
\label{fig2}
\end{figure}

\begin{figure}[htb]
\epsfxsize=3.0 in
\bigskip
\centering
        \begin{minipage}[b]{0.5\textwidth}
        \centering
            \subfigure{
              \includegraphics[width = 2.8 in]{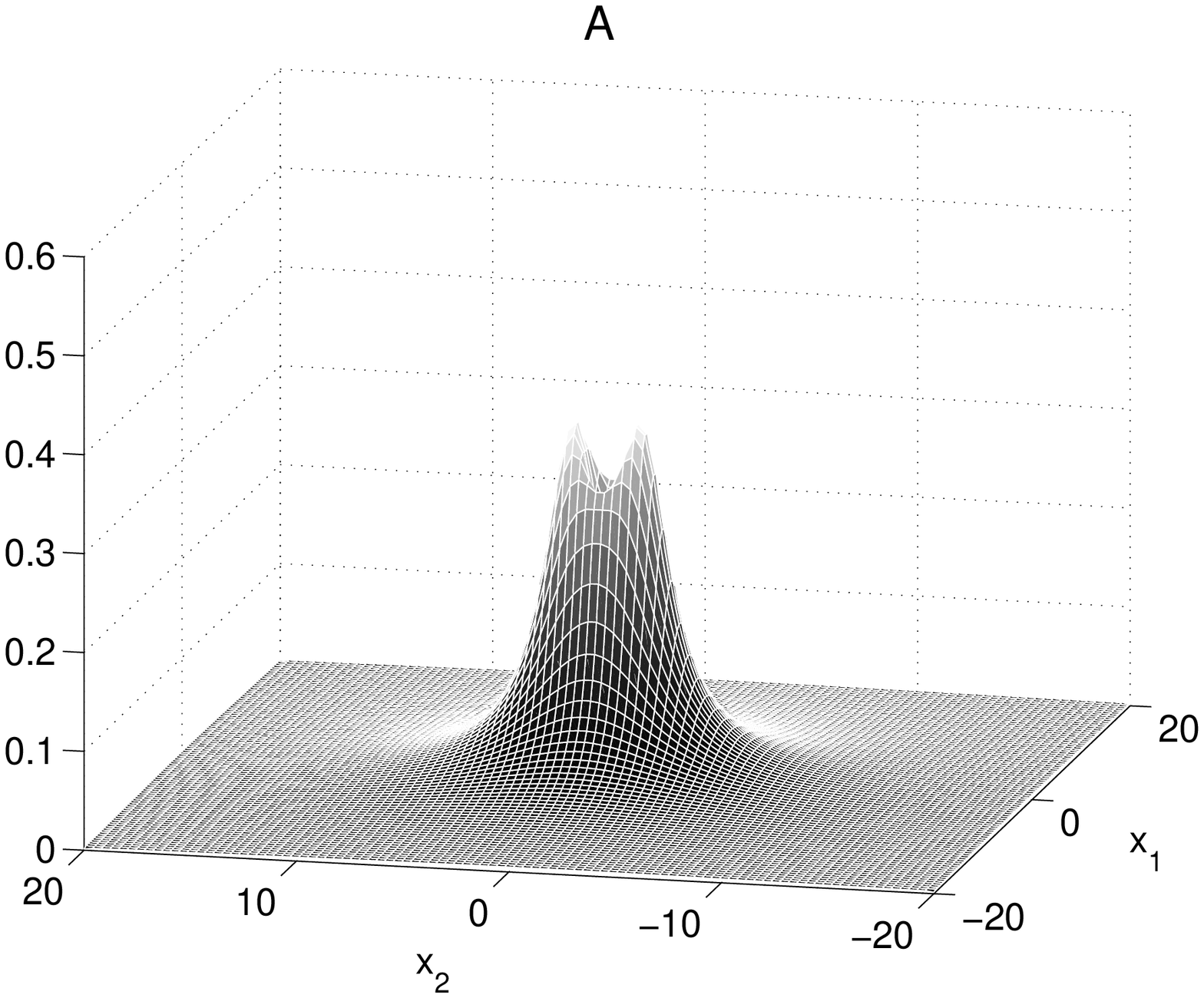}}
        \end{minipage}%
        \begin{minipage}[b]{0.5\textwidth}
        \centering
            \subfigure{
             \includegraphics[width = 2.8 in]{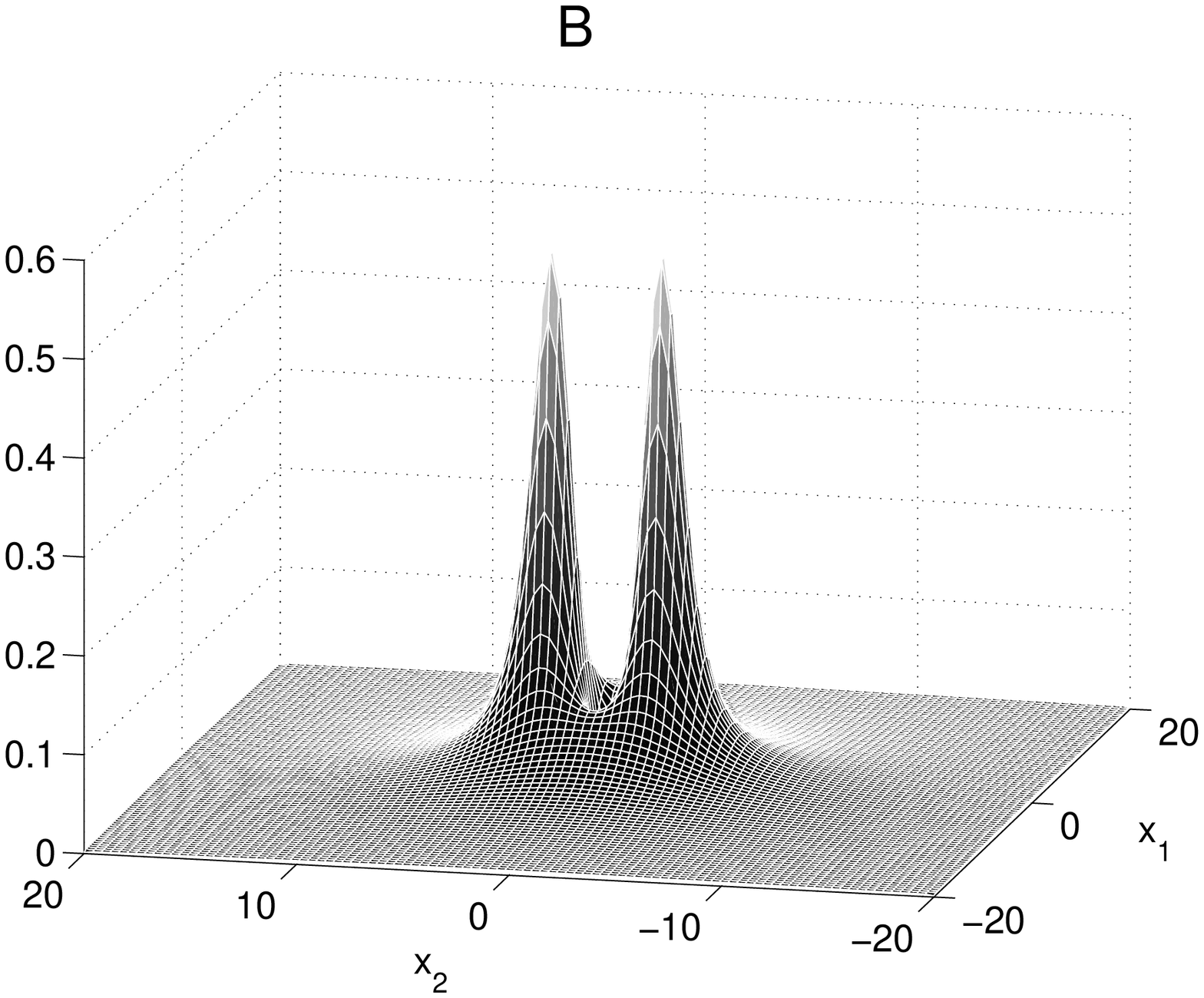}}
        \end{minipage}
\caption{\baselineskip=12pt {\it
Energy density  in the $x_1x_2$-plane of the 
          vortex configurations A and B of figure 2.}
}
\bigskip
\label{fig3}\end{figure}

In our numerical simulation of the gradient flow we 
generated 700 configurations.  The most convenient way
of labelling these configurations is in terms of  
 their Ginzburg-Landau energy rather than by the configuration 
number or the 
gradient flow ``time'' $\tau$, which have no physical significance
in the Gross-Pitaevski model.
The finite lattice used in the computation provides a natural
renormalisation of the energy of the vortex configurations, assigning the
value $E_{\mbox{\tiny ren}}[\psi_2] =V_{\mbox{\tiny max}}=31.982$ to
 the saddle point configuration $\psi_2$. The configuration
A in figures 2 and 3 has  the  energy $V=31.955$  
and the configuration B has the energy $V=30.257$.

There is another way of parametrising
the gradient flow which we need to address, namely the 
parametrisation in terms
of vortex separation.  Vortex separation is
used as a parameter for pinned vortex configurations in
the work of Ovchinnikov and Sigal.
In view of our later comparison with that work we therefore  
make a few comments on it here. The pictures of the vortex
configurations and energy densities shown in figures 2 and 3
illustrate  that  the notion of a separation
is not well-defined during the early stages of the gradient
flow. The best one can do is to define a functional of vortex
configurations  which reproduces the distance between the 
maxima of the energy density (or, equivalently, between the
zeros of the  field $\psi$) for well-separated vortices
and which is computationally convenient. 
In the following we use the vorticity  (\ref{vorticity})
instead of the energy density.
The vorticity looks qualitatively very similar to the energy
density during all stages of the gradient flow and in particular
it is strongly peaked around the zeros of the field $\psi$
for well-separated vortices. The advantage of using the vorticity rather
than the energy density 
is that the total vorticity is conserved whereas the total energy
changes during the gradient flow.
To motivate our approach further,  note  that the configurations
generated during the gradient flow
break the invariance group $R_2$ of the saddle point
  to the  group $\ZZ_2\times \ZZ_2$ generated by
 reflections about the $x_1$ and $x_2$ axes, each combined with
the complex   conjugation of $\psi$. 
As a result the vorticity $\gamma$  of all configurations is 
reflection symmetric about two orthogonal axes.
 For our gradient flow the reflection
axes are the coordinate axes,  and the vortices separate
along  the $x_2$-axis because we 
have broken the rotational symmetry $R_2$ of the saddle point 
solution explicitly by using the perturbation $\xi$ (\ref{eigenfun}).
A different perturbation would have led to an isomorphic reflection 
symmetry about a different set of orthogonal axes,
and a different separation direction. 
We use the reflection axis orthogonal to the separation direction
to divide $\RR^2$ into
two half spaces $H_r=\{(x_1,x_2)\in\RR^2|x_2\geq 0\}$ and 
$H_l=\{(x_1,x_2)\in\RR^2|x_2 < 0\}$.  
Then we define the separation functional as 
\bea
\label{distdef}
a[\psi]= \int_{H_r} x_2\,\gamma \, d^2x
-\int_{H_l} x_2\,\gamma \, d^2x= 2\int_{H_r} x_2\,\gamma \, d^2x .
\eea
 As expected, the 
separation  functional (\ref{distdef})  increases monotonically 
during the gradient flow, but one interesting feature is that 
it assigns a non-vanishing value to the ``coincident'' vortex
configuration $\psi_2$. This  is a familiar feature
of separation  parameters used for
other solitons, such as monopoles in non-abelian gauge theory
\cite{AH}.
The point is that solitons lose their separate identity
when they overlap and separation ceases to be a meaningful
concept. In particular there is no reason 
to insist that a separation  parameter vanish for a particular
configuration.
Numerically, we find the minimal separation to 
be $a_{\mbox{\tiny min}}=a[\psi_2]=4.815$, 
which corresponds roughly to 
the diameter of the ring on which the energy density of $\psi_2$ 
is maximal. 
For the  configuration A in figures 2 and 3 
 we find $a=4.896$ and for configuration B we find $a=6.406$.  The lowest
energy configuration we generated has the energy $V=24.405$ and 
separation $a=15.570$.

For the construction of the unstable manifold
we return to the parametrisation of  the gradient flow curve 
in terms of renormalised energy. Thus we obtain  a family of fields
$\hat \Psi(V;x)$,
labelled by the value $V$ of the renormalised Ginzburg-Landau energy.
According to the general prescription of \cite{Manton1} we should now
act with the full symmetry group (\ref{symgroup}) to generate a family
of saddle points and a family of gradient flow curves, and use their
union as the collective coordinate manifold for the truncated dynamics.
However, we are only  interested in the relative motion of the 
two vortices in the fission process and not in  their centre-of-mass motion.
The latter can be found  by applying Galilean boosts (\ref{galileo})
to the entire
configuration. Furthermore, we exclude collective coordinates which
change our boundary condition. Both phase rotations and spatial rotations
 change the field ``at infinity'' (or at the boundary of our lattice),
but the subgroup $R_2$ defined in (\ref{sadsym}) respects the boundary
condition. It also leaves the saddle point solution $\psi_2$ invariant
but it acts non-trivially on the gradient curve emanating from $\psi_2$.
Thus we obtain a family of fields
\bea
\label{thefamily}
\Psi(V,\chi;r,\theta)= \hat\Psi(V;r,\theta-\chi)e^{2i\chi}
\eea
depending on two collective coordinates $V,\chi$. The former measures
the energy and the second labels the spatial orientation of
 the two-vortex configurations generated during the gradient flow.
Since the saddle point configuration with energy $V_{\mbox{\tiny max}}$
does not depend on the angle $\chi$, the collective coordinate
manifold is topologically a plane, with $(V_{\mbox{\tiny max}}-V,\chi)$ as 
polar coordinates. We denote this manifold by  ${\cal M}_2$.
The range of the angle $\chi$ is $[0,\pi)$ since the configurations
$\hat \Psi(V;x)$ generated during the gradient flow are invariant under
a rotation by $\pi$ (which is the product of the reflections at the 
coordinate axes discussed above).
Note that ${\cal M}_2\subset {\cal C}_2$.

It is now a simple matter to compute the restriction of the 
Gross-Pitaevski  Lagrangian  (\ref{GPlag})  to 
${\cal M}_2$. We allow the collective coordinates $V$ and $\chi$ 
to depend on time and insert the fields $\Psi(V(t),\chi(t);r,\theta)$
into  (\ref{GPlag}). The result is 
\bea
\label{indlag}
L[\Psi(V,\chi;\cdot)]=
- \frac 12 \bigl( 2Q(V)-J(V)\bigr) \dot\chi  - \frac 12G(V)\dot V - V ,
\eea
where we have written $ Q(V)$ and  $J(V)$ 
for  the Noether charges (\ref{qcharge}) and 
(\ref{jcharge}) evaluated on $\Psi(V,\chi;x)$, i.e.
\bea
Q(V)=\int |\Psi(V,\chi;x)|^2 \, d^2x
\eea
and
\bea
J(V)=\int \mbox{Im}\left(
\bar{\Psi}(V,\chi;x)\frac{\partial \Psi}{\partial \theta}
(V,\chi;x)\right) \, d^2x.
\eea  
The integrals defining  $Q$ and $V$ are both divergent, but the combination
\bea 
\label{Kfunctional}
K(V)= J(V)-2Q(V)
\eea 
occuring in the Lagrangian is finite. To see this note that for 
the saddle point configuration $\psi_2$ one computes  $J=2Q$,
 so that $K(V_{\mbox{\tiny max}})=0$.
As the configuration evolves away from the saddle point $K$ grows
but it remains finite for any finite value of $V$. This follows
from the continuity of the semigroup defined by the evolution
equation \ref{heat} in a suitable norm, see \cite{GV1} and \cite{GV2},
and from the continuity of the funtionals $J$ and $Q$ (\ref{qcharge}) and 
(\ref{jcharge}) with respect to that norm.
Finally, the function $G(V)$ is 
\bea
\label{Gcharge}
     G(V)=\int \mbox{Im}\left(
\bar{\Psi}(V,\chi;x)\frac {\partial \Psi}{\partial V}(V,\chi;x)\right)\, d^2x.
\eea
Neither $K$ nor $G$  depends on $\chi$
because of the rotationally invariant integration measure and 
 because the integrands
are  independent of the phase of $\Psi$.  
The term  
$G(V)\dot V$ is a total time derivative which does not affect the 
equations of motion. We therefore omit it in
 our final expression for the induced Lagrangian $L_2$ on ${\cal M}_2$:
\bea
L_2 = \frac 12 K(V) \dot\chi -V. 
\eea
The Euler-Lagrange equations are very simple because our 
truncated system, like the original Gross-Pitaevski dynamics, is
rotationally invariant and has a conserved energy. Thus $V$ remains
constant during the time evolution, and the angle $\chi$ changes 
according to 
\bea
\label{chidot}
\dot \chi(V) = \frac 2 {K'(V)}.
\eea
It follows from the constancy of $V$ that $\dot\chi$ is also constant
during the time evolution. We introduce the abbreviation
\bea
\omega(V) = \dot \chi(V)
\eea
for the rate of change of the angle $\chi$. The dependence of this
angular velocity on $V$ is the main dynamical information we extract
from our truncated dynamics. 
It is is given by the differential quotient
\bea
\label{angvel}
\omega(V) = 2 \frac {d V}{dK}.
\eea
Numerically, we compute $\omega$ by evaluating  finite
difference quotients $\Delta V/\Delta K$ of successive configurations.
Our results are displayed in figure 4.
In the early phase of the gradient flow (roughly the 
first 100 configurations)  both $ V$ and $ K$
change very slowly and 
accurate numerical computation of the difference quotient is difficult.
We have omitted  these difference quotients from our plot. Since
the energy $V$ hardly changes during this part of the gradient flow,
this omission makes no visible to  difference to the graph $\omega(V)$ 
shown in  figure 4. The next section is devoted to a detailed discussion
and interpretation of our results.

\begin{figure}[htb]
\epsfxsize=3 in
\bigskip
\centerline{\epsffile{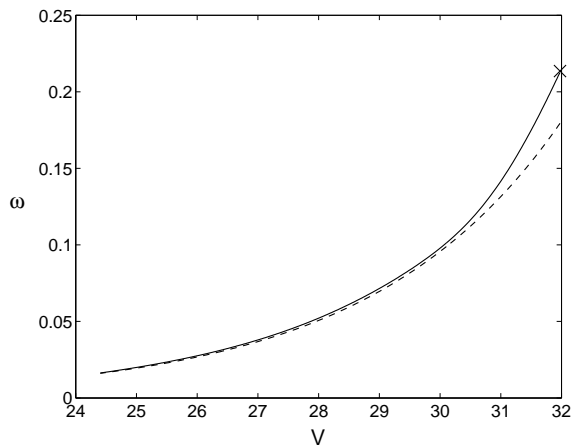}}
\caption{\baselineskip=12pt {\it
The solid line shows 
the angular velocity $\omega$ of two-vortex 
configurations as a function of the energy $V$.
The dependence of $\omega$ on the energy predicted 
by the Kirchhoff-Onsager law is plotted with a dashed line. 
The value of $\omega_{\mbox{\tiny max}}$ given in  
(\ref{omval}) is marked with a ``$\times$''. } }
\bigskip
\label{fig4}
\end{figure}

\section{Orbiting vortex pairs}

Our two-dimensional Lagrangian dynamical system $({\cal M}_2,L_2)$ 
is designed to approximate the slow interactive dynamics of two
vortices moving according to the Gross-Pitaevski equation (\ref{GPequation}).
The model predicts that the two vortices will orbit each other, with
the angular frequency depending on their total energy as shown in our
figure 4. While the qualitative behaviour of the two vortex system
is not surprising and familiar from vortices in ordinary fluids and 
also vortices in the gauged Ginzburg-Landau model \cite{Manton2},
the precise prediction of the variation 
of the angular frequency with the energy for both overlapping and 
well-separated vortices is new. In \cite{Dziarmaga} 
Dziarmaga computed the rotation
frequency of an overlapping  vortex pair by linearising the Gross-Pitaevski
equation around the saddle point solution $\psi$. The linearised
solution is 
\bea
\label{linsol}
\psi (t,x) = \psi_2(x) + \epsilon e^{-i\lambda t} (u(r) + e^{4i\theta}v(r)),
\eea  
where $\lambda$ and $(u,v)$  are the eigenvalue and 
eigenfunctions of the eigenvalue problem (\ref{coupled}). The interpretation
of the  linearised solution in terms of rotating vortices requires some 
care. As we saw in section 3, the radial eigenfunction $v$ is much smaller
than $u$. Thus, as a first approximation, we may think of (\ref{linsol})
as a field whose magnitude depends on $r$ according to $\epsilon u(r)$ 
and whose
direction changes with angular frequency $\lambda$, all  superimposed on the 
static $n=2$ solution. The effect of  such a superposition is a configuration
with two zeros (whose separation depends on $\epsilon$) which rotate around 
one another with angular velocity $\omega_{\mbox{\tiny lin}} = -\lambda/2$.
The factor $1/2$ arises  because for $\psi_2$ a phase rotation by $\alpha$
is equivalent to a spatial rotation by $\alpha/2$. With our value for 
$\lambda$, we thus find 
$ \omega_{\mbox{\tiny lin}} = -\lambda / 2 = 0.20935$.

In the limit 
$V \rightarrow V_{\mbox{\tiny max}}$
our numerically computed  function $\omega(V)$ 
 approaches a value which is close to 
but not exactly equal to $\omega_{\mbox{\tiny lin}}$. 
The discrepancy arises because 
 we neglected the radial eigenfunction $v$ in our interpretation of the 
linearised solution. We can establish
 a  precise relation between the eigenvalue  $\lambda$ and the 
limiting angular frequency
\bea
\label{maxfre}
\omega_{\mbox{\tiny max}}:=\lim_{V\rightarrow  V_{\mbox{\tiny max}}}
\omega(V)
\eea 
as follows. Using the definition (\ref{Kfunctional}) and 
noting  that both $V$ and $K$ are originally defined along the 
gradient flow curve parametrised in terms of the auxiliary
variable $\tau$ (\ref{heat}) we write
\bea 
\omega(V) = 2 \frac{d V}{dK}=2 \frac{dV}{d \tau}/\frac{dK}{d\tau}.
\eea
 Recall that  we computed our gradient 
flow curve by starting at $\tau=0$ with the configuration $\psi_2 +\epsilon
\xi$, where $\xi$ is the eigenmode (\ref{eigenfun}) and $\epsilon$
is small (in practice $\epsilon =0.0005$). 
At that point the gradient of the Ginzburg-Landau 
energy functional is 
\bea
\frac {\delta E}{\delta\bar{ \psi}}[\psi_2+\epsilon\xi]& \approx& 
\epsilon \frac{\delta^2 E}{\delta \psi\delta \bar{\psi}}\xi +
\epsilon \frac{\delta^2 E}{\delta \bar{\psi}^2}\bar\xi 
\nonumber \\
&=&\frac 1 2\lambda\epsilon\xi,
\eea
where we have used that the first functional derivative of $E$
vanishes at $\psi_2$ and that $\xi$ is an eigenfunction
of the Hessian, i.e. it satisfies (\ref{hesseigen}).
Hence from (\ref{heat})
\bea
\label{useful}
\partial_\tau\psi_{|_{\tau=0}}\approx -\lambda\epsilon \xi
\quad \mbox{and} \quad
\partial_\tau\bar{\psi}_{|_{\tau=0}} \approx -\lambda\epsilon \bar{\xi}
\eea
and thus
\bea
\left.\frac{dV}{d \tau}\right|_{\tau=0} &=& \int
\frac {\delta E}{\delta \psi}[\psi_2+\epsilon\xi]\,\,
\partial_\tau\psi_{|_{\tau=0}}+
\frac {\delta E}{\delta \bar{\psi}}[\psi_2+\epsilon\xi]\,\,
\partial_\tau\bar{\psi}_{|_{\tau=0}} 
\,d^2x \nonumber \\
&\approx& -\epsilon^2\lambda^2\int \bar{\xi}\xi \,d^2x \nonumber \\
&=&-\epsilon^2\lambda^2 \cdot2 \pi 
\int (u^2 +v^2)\,rdr,
\eea
where we used  (\ref{useful}) and 
inserted  the expression (\ref{eigenfun}) for $\xi$.

It is easy to check that any $R_2$-invariant configuration (\ref{sadsym})
and hence in particular  $\psi_2$ is a stationary point of the functional
 $K[\psi]$  (\ref{Kfunctional}). Thus  $\frac{\delta K}
{\delta \psi} (\psi_2) =0$
and by a similar calculation to the one given above we have  
\bea
\left.\frac{dK}{d\tau}\right|_{\tau=0} &=& \int
\frac {\delta K}{\delta \psi}[\psi_2+\epsilon\xi] 
\,\,\partial_\tau\psi_{|_{\tau=0}}
+\frac {\delta K}{\delta \bar{\psi}}[\psi_2+\epsilon\xi] 
\,\,\partial_\tau\bar{\psi}_{|_{\tau=0}}
\, d^2x \nonumber \\
 &\approx&-\epsilon^2\lambda
\int \bar{\xi} \frac {\delta^2 K}{\delta \bar{\psi}\delta \psi}\xi
+\xi\frac {\delta^2 K}{\delta \psi\delta \bar{\psi}}\bar{\xi}
\nonumber \\
 &=&-2\epsilon^2\lambda
\int \left(\mbox{Im}\bar{\xi}\partial_\theta\xi -2\xi\bar{\xi}\right)\,
 d^2x
\nonumber \\ 
&=& 4 \epsilon^2\lambda \cdot 2\pi 
\int (u^2 -v^2)\,r dr,
\eea
where we used that $\delta^2K/\delta \psi^2=\delta^2K/\delta \bar{\psi}^2=0$.
The approximate equalities become exact in the limit $\epsilon
\rightarrow 0$.
Thus we find the following exact expression for the angular
frequency $\omega_{\mbox{\tiny max}}$  (\ref{maxfre}) in 
 terms of the 
eigenvalue $\lambda$:
\bea
\label{goodformula}
\omega_{\mbox{\tiny max}} = -\frac \lambda 2
\frac{ \int  \, ( u ^2  +  v^2)rdr}{ \int \, (u^2- v^2)rdr}.
\eea
In view of the 
numerical difficulties in computing $\omega$ near the saddle
point this formula is very useful in practice.
Evaluating it    numerically we find
\bea
\label{omval}
\omega_{\mbox{\tiny max}} = -\frac \lambda 2 \times  1.0200 = 0.21354.
\eea
This value provides a check on our computation of 
$\omega(V)$ near the saddle point 
 and  is marked with a ``$\times$'' in figures 4 and 5. 

For well-separated vortices we can compare our results with
those obtained    by Neu in  \cite{Neu1,Neu2}  via scaling techniques 
 or those obtained by Ovchinnikov
and Sigal  in \cite{OS1} based on 
 pinned vortex configurations.  Both find that
well-separated vortices are governed by the Kirchhoff-Onsager
law.  According to that law  two vortices separated by a distance
$a$  have an interaction energy given by
\bea
\label{endist}
V(a) = -2\pi \ln  a  + C,
\eea
where $C$ is an arbitrary normalisation constant, 
and orbit each other with angular frequency
\bea
\label{simplelaw}
\omega_{\mbox{\tiny KO}} = \frac{4}{a^2}. 
\eea
It follows that the rotation frequency is given as a function
of the energy via
\bea
\label{KOlaw}
\omega_{\mbox{\tiny KO}} (V) = \tilde C e^{\frac V \pi},
\eea
where $\tilde C= 4 \exp(-C/\pi)$.
The constant $C$ (and hence $\tilde C$) is determined once
the dependence of the energy on the vortex separation is known.
To make contact with the Kirchhoff-Onsager relations we thus
have to make use of the separation functional introduced for
our vortex configurations in (\ref{distdef}). We pick a
vortex configuration  consisting of clearly separated vortices
and compute its energy to be $V=27.253$ and its separation 
to be $a=10.012$. Inserting these values into (\ref{endist})
fixes the value of $C$ and hence $\tilde C$. We have plotted
the resulting function (\ref{KOlaw}) in figure 4. 
It is important to be clear about the interpretation of the 
two curves in figure 4. For well-separated vortices
we can meaningfully associate a separation parameter to a vortex
configuration. The Kirchhoff-Onsager relation
 gives the angular frequency of point-vortices with that  
separation. The fact that 
for low energy configurations consisting of well-separated
vortices the Kirchhoff-Onsager relation agrees with our 
results shows that well-separated vortices may be treated
as point-vortices, in agreement with the results of 
 Ovchinnikov and Sigal in \cite{OS1}. 
The two curves in figure 5 differ for 
configurations near the saddle point, not so much 
because the predictions of our model disagree
with those of the point vortex model but because
the two models are no longer compatible in that regime.

From the point of view of the Kirchhoff-Onsager relation,
the dependence (\ref{simplelaw}) of the angular velocity
 on separation is more fundamental than the dependence on the energy. 
In figure 5 we  plot the angular velocity $\omega$ as a function
of the separation parameter (\ref{distdef}) and compare it
with the simple Kirchhoff-Onsager prediction. Again the 
two plots agree for large separations, and the comments 
made above  on the comparison near the saddle point apply 
here, too. However, it is worth emphasising that the 
Kirchhoff-Onsager relation predicts a divergence of the 
angular velocity 
when the separation tends to zero. 
Our results show that  for the  Ginzburg-Landau
vortices this divergence is removed, essentially  by cutting off 
small separation parameters. This is yet another  example of 
a soliton model ``regularising'' the singularities of a 
point-particle model.

\begin{figure}[hbt]
\epsfxsize=3 in
\bigskip
\centerline{\epsffile{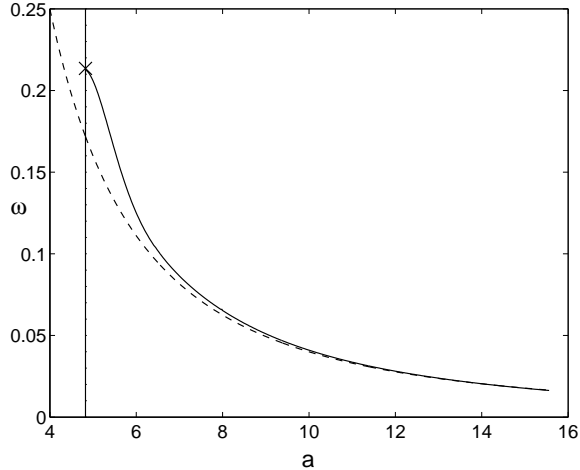}}
\caption{\baselineskip=12pt {\it
The solid line shows the  angular velocity $\omega$ of two-vortex 
configurations as a function of the separation $a$.
The dependence of $\omega$ on the separation  predicted 
by the Kirchhoff-Onsager law is plotted with a dashed line. 
The value of $\omega_{\mbox{\tiny max}}$ given in (\ref{omval}) 
is marked with a ``$\times$''.}}
\bigskip
\label{fig5}
\end{figure}

\section{Discussion and Conclusion}

The unstable manifold method used in this paper to approximate
the dynamics of interacting vortices in the Gross-Pitaevski model
allows one to analyse the relative motion of  two vortices at 
arbitrary separation. When the vortices are overlapping our results
reproduce those found with linearisation methods and  when the vortices
are well-separated our results agree with the Kirchhoff-Onsager
law. Our method gives a unified description of both regimes,
and successfully interpolates between them.

When vortices evolve according to the Gross-Pitaevski equation
they will in general also emit radiation. This effect is not captured
in our approximation. However, it is possible to compute radiative
corrections to the vortex motion predicted by our approximation
using the methods developed by Ovchinnikov and Sigal in \cite{OS2}. 
In particular, Ovchinnikov and Sigal studied the radiative corrections
 to the orbiting motion of a vortex pair and found that, for large
vortex separation $a$, the power emitted in the radiation is proportional
to $a^{-6}$. As a result of this energy loss, the orbiting vortices
 drift apart slowly as they orbit each other. However, since the power
loss is  small, the change of the separation during one period
of rotation is also small (much less than 1 \%). 
Combining these results
with our calculations, we arrive at the following qualitative picture
of vortex fission in the Gross-Pitaevski model. Suppose the $n=2$ 
saddle point solution is perturbed and allowed to evolve according
to the Gross-Pitaevski equation. The  perturbed vortex configuration
 rotates and emits radiation. The radiation carries away energy 
and the vortex configuration adjusts by finding a configuration
of lower energy. The most efficient way of doing  so is by
following a path of steepest (energy) descent. As we have seen, the 
$n=2$ vortex breaks up into two $n=1$ vortices along the path of 
steepest descent. Thus we can use our truncated dynamical system
$({\cal M}_2,L_2)$ and expect the vortex pair to orbit with 
the angular frequency depending on the energy according to (\ref{chidot}).
The continued rotation results in further emission of radiation
and loss of energy, to which the vortex pair adjusts by drifting further
apart. The fission process is thus  governed by the combination of 
two  effects:  the radiation
carries away energy and the vortex configuration adjusts by sliding 
down the unstable manifold of the $n=2$ saddle point. 
It would be interesting to make this picture more precise by
studying the radiation effects quantitatively. This should 
probably be done in conjunction with a  more careful study of the conservation
laws in the theory, similar to the investigation 
for gauged vortices in \cite{MN}.  As mentioned briefly in section 2, the 
conservation of angular momentum can also be expressed as the 
conservation of the moment (\ref{Jmoment}) of the vorticity. 
Just looking at vortex motion during the fission process it seems
 that the moment (\ref{Jmoment}) increases as the vortices drift
apart. It would be interesting to understand how 
 the radiation makes up for this change, possibly by
carrying  negative  vorticity  off  to infinity.

We would like to  emphasise that the work described here
is the first application of the 
unstable manifold method to  soliton dynamics in more than one
spatial dimension. As such, it contains a number of useful general
lessons. At first sight it may not seem sensible to   use one non-linear 
 partial differential
equation (the gradient flow equation) to approximate another non-linear
partial differential equation (the Gross-Pitaevski equation).
However, generally speaking,  gradient flow is easier to understand
and to compute  numerically than Hamiltonian flow like that defined by
the energy-conserving Gross-Pitaevski equation. One  reason for
this is that the typical time scales of the two evolution  processes 
are very different. 
Our discussion of the vortex fission process illustrates 
this point. The solution of the gradient flow equation over a relatively 
short CPU time (roughly 700 steps) maps out all the interesting configurations
 between the $n=2$ saddle point and the well-separated vortices.
The fission process  according to the Gross-Pitaevski equation, by contrast,
is expected to take several orders of magnitude longer, making it difficult
to maintain numerical accuracy.  The gradient flow
equation provides an efficient way of  mapping out 
those configurations which are relevant for slow (or low-energy)  dynamics. 
At a more practical level, we found that the explicit study of the linear
negative modes of the saddle point was essential for an efficient
 computation of  the unstable manifold. We expect that this will become
even more important for saddle points with more than one negative mode.
Thus one could repeat the calculations performed here for the saddle
point solutions $\psi_n$ with vortex number $n>2$. As explained in 
\cite{OS3},  there is now more than one unstable mode. As a first 
step in constructing the unstable manifold of these saddle points one
needs to  compute the negative eigenmodes of the linearised Ginzburg-Landau 
equation  explicitly.  While the explicit construction of the unstable
manifolds and the computation of the induced dynamics may be 
difficult, it would be interesting to understand general geometrical
features of the unstable manifolds and use them to derive qualitative
features of multi-vortex dynamics in the Gross-Pitaevski model.

It remains an open   problem to justify
  the unstable manifold
approximation used in this paper analytically. The fact that our 
results  agree with those obtained via  
more familiar approximations in limiting regimes 
 provides an encouraging check. However, this agreement also
raises a question. Near the saddle point  our method
 agrees with results obtained by  linearisation, and  the small
  parameter controlling the  validity
of the  approximation is the amplitude $\epsilon$ in (\ref{linsol}).
As explained, this is related to the separation of the two zeros
of the field $\psi$. Far away from the saddle point, our approximation
agrees with the point-vortex model implicit in the Kirchhoff-Onsager
  law. The small parameter controlling the validity of the
  approximation in this regime is the {\it inverse} separation of the 
zeros of $\psi$ (i.e. the inverse distance between the vortices). Since our 
method interpolates between the two approximations it is not clear
which, if any, small parameter controls the approximation in 
the intermediate regime.
More generally, proof of the validity of the 
approximation, possibly  along the lines of the proof of the geodesic
approximation for gauged vortex dynamics in \cite{Stuart}, would 
be very desirable. A useful starting point for such an investigation
may be the following natural relation between the gradient flow
equation (\ref{heat}) and the Gross-Pitaevski equation (\ref{GPequation}).
Both define flows in the space ${\cal C}$ (\ref{confspace})
 in terms of 
the  Ginzburg-Landau energy. However,  the gradient flow uses 
the  inner product (\ref{metric}) whereas  the Gross-Pitaevski equation
is a Hamiltonian flow using the symplectic structure (\ref{symform}).
The inner product (\ref{metric}) and the symplectic form (\ref{symform})
are equal  to, respectively,
 the real and imaginary part of the sesquilinear
form
\bea
g(\vec{\xi},\vec{\eta})=\int_{\RR^2}\xi \bar{\eta}\,\, d^2x.
\eea
One interesting manifestation of this relationship is that 
gradient flow  trajectories and solutions of the Gross-Pitaevski
equation are orthogonal with respect to the inner product (\ref{metric})
whenever they intersect.

\noindent {\bf Acknowledgements}

\noindent Most of the research reported here was carried out while
OL  was an MSc student at Heriot-Watt University. 
OL thanks the  DAAD for a scholarship during that  time
and  Dugald
Duncan for advice on the numerical calculations.
BJS acknowledges an EPSRC advanced research fellowship.


\begin{thebibliography}{99}
\bibitem{BBH}
Bethuel F, Br\'ezis H  and H\'elein F  1994 {\em Ginzburg-Landau vortices}
(Basel: Birkh\"auser)
\bibitem{OS3} Ovchinnikov Y N and Sigal I M  1997
Ginzburg-Landau equation I. Static vortices,  in
 Partial differential equations and their applications
{\em CRM Proceedings and Lecture Notes} {\bf 12} 
199--220, eds Greiner P et al (Providence: American Mathematical Society) 
\bibitem{Neu1}
Neu J C  1990  Vortices in complex scalar fields,
{\em Physica} {\bf D 43} 385--406
\bibitem{Neu2} Neu J C  1990  Vortex dynamics of 
the non-linear wave equation {\em  Physica} {\bf D 43}  407--420
\bibitem{LX1}
Lin F-H and Xin J X  1999 On the imcompressible fluid limit and 
the vortex motion law of the nonlinear Schr\"odinger equation
{\em Commun.~Math.~Phys.}  {\bf 200}  249--274 
\bibitem{LX2}
Lin F-H and Xin J X  1999  On the dynamical
law of the Ginzburg-Landau vortices on the plane
{\em Comm.~Pure~Appl.~Math.} {\bf 52}  1189--1212
\bibitem{CJ}
Colliander J E and  Jerrard R L 1998 Vortex dynamics for the 
Ginzburg-Landau-Schr\"odinger equation {\em  Internat.~Math.~Res.~Notices}
{\bf 7} 333--358
\bibitem{OS1}
Ovchinnikov Y N and Sigal I M 1998 The Ginzburg-Landau equation III.
Vortex dynamics {\em  Nonlinearity} {\bf 11} 1277--1294
\bibitem{OS2}
Ovchinnikov Y N  and  Sigal I M
1998  Long-time behaviour of Ginzburg-Landau vortices
{\em Nonlinearity} {\bf 11} 1275--1310
\bibitem{Manton1}Manton N S 1988  Unstable manifolds and soliton dynamics 
{\em Phys.~Rev.~Lett.} {\bf 60} 1916--1919
\bibitem{MM} Manton N S and  Merabet H 1997  $\phi^4$ kinks -- gradient flow
and dynamics {\em  Nonlinearity} {\bf 10} 3--18    
\bibitem{PT}
Papanicolaou N and  Tomaras T N  1993 On the dynamics of 
vortices in a nonrelativistic Ginzburg-Landau model {\em Phys.~Lett.} 
{\bf A 179}  33
\bibitem{Dziarmaga} Dziarmaga  J 1996  Dynamics of overlapping vortices
in complex scalar fields {\em  Act.~Phys.~Pol.} {\bf B27} 
1943--1960
\bibitem{GV1}Ginibre J  and  Velo G   1996  The Cauchy problem in 
local spaces for the complex Ginzburg-Landau equation I. Compactness
methods {\em  Physica}  {\bf D 95} 191--228
\bibitem{GV2} Ginibre J  and  Velo G 1997  The Cauchy problem in 
local spaces for the complex Ginzburg-Landau equation II.
Contraction methods {\em  Commun.~Math.~Phys.} {\bf 187}  45--79
\bibitem{Manton2} Manton N S 1997  First order vortex dynamics
{\em  Annals Phys.} {\bf 256}  114--131
\bibitem{MN}
Manton N S  and  Nasir S M 1999
Conservation laws in a first order dynamical system of vortices
{\em Nonlinearity}  {\bf 12}  851--865
\bibitem{AH} Atiyah M  and Hitchin N 1988 {\em Geometry and dynamics
of monopoles} (Princeton: Princeton University Press)
\bibitem{Stuart} Stuart D 1994  Dynamics of
abelian Higgs vortices in the near Bogomolny regime
{\em Commun.~Math.~Phys. }{\bf 159}  51--91.
\end{thebibliography}
\end{document}